\documentclass[epj]{svjour}
\usepackage{graphics}
\usepackage{amsmath}
\usepackage{amssymb}
\usepackage{amsfonts}
\usepackage[T1]{fontenc}
\usepackage{bm}
\usepackage{mathptmx}
\usepackage{color}
\usepackage{slashbox}
\begin{document}
\title{Imitation dynamics in a game of traffic}

\author{G.H. Paissan\inst{1} \and G. Abramson\inst{2}
\thanks{\emph{address:paissan@cab.cnea.gov.ar} }%
}                     
\offprints{Gabriel H. Paissan}          
\institute{Consejo Nacional de Investigaciones Cient\'ificas y T\'ecnicas, Centro
At{\'{o}}mico Bariloche and Centro Regional Universitario Bariloche - UNCo, 8400
Bariloche, R\'io Negro, Argentina \and Consejo Nacional de Investigaciones
Cient\'ificas y T\'ecnicas, Centro At{\'{o}}mico Bariloche and Instituto Balseiro,
8400 Bariloche, R\'io Negro, Argentina}
\date{Received: date / Revised version: date}
%
\abstract{We study a model of traffic where drivers adopt different behavioral strategies.
These can be cooperative or defective according to a driver abiding or not by a
traffic rule. Drivers can change their strategy by imitating the majority, with a
rule that depends on the strategies with which they have interacted.
These interactions occur at intersections, where vehicles pay a temporal cost according
to their strategy. We analyze the conditions under which different strategy compositions represent an advantage in the system velocity.  We found that the cooperators' mean
speed is higher than the defectors' even when the vehicle density is large. However,
defectors can obtain benefits in their mean speed when they are a minority in an
essentially cooperative population. The presence of a core of educated drivers, who
persist firmly in a cooperative behavior, optimizes the speed in the system,
especially for intermediate values of vehicular density and higher temporal costs.
} 
\maketitle
\section{Introduction}

\label{intro}Vehicular traffic dynamics has received considerable attention since, at least,
the middle of the twentieth century. In the decades that followed different points
of view have been used to tackle different features of the many problems associated
with traffic. The interest of these hardly needs justification: the many aspects of
traffic have enormous impact in our civilization, with applications in many fields,
ranging from engineering to the social sciences. The volume of traffic flow has
quickly overpassed the capacity of the cities and highways, once and again, country
after country, making the understanding of its dynamic an imperative in many
societies. The problems of infrastructure and urban development able to accommodate
an increasing flow range reside at one end of the spectrum. At the other end there are
aspects of education, social planning and law enforcement, aiming at helping the
flow in the best interest of the society. Many other ancillary problems lay in
between: congestion, pollution management (including noise and vibration),
optimization of energetic resources, economics, reduction of accidents and
casualties, parking, public transportation alternatives, remote monitoring,
robotization, etc.

The complex spectrum of mathematical traffic models is well documented in the
reviews by Helbing and Chowdhury \cite{helbing2001,chowdhuryetal2000}.  The
influential works of Lighthill and Whitham \cite{lighthill1955}, and then of
Richards \cite{richards1956}, are based on a macroscopic approach in which traffic
is considered a continuous medium---like a fluid---characterized by a spatial
density and a velocity field. On a different vein, ``microscopic'' models that treat
the motion of each vehicle separately have been developed as a parallel line of
research. Microscopic models emphasize the role played by the (non linear)
interactions between vehicles. The interest of physicists in traffic problems
received an enormous boost with the work of Nagel and Schreckenberg
\cite{nagel1992}. There exist very detailed microscopic models that serve the
purpose of analyzing mostly local situations. Such level of detail is still
prohibitive in the study of large scale features of traffic flow.
Nagel-Schreckenberg models and others, on the other hand, exploit the more
manageable representation of traffic as cellular automata.

In this work we present a bidimensional model of traffic where drivers operate with different behavioral
strategies. In this framework we introduce a dynamic of imitation based on the types
of strategies faced by drivers at each intersection of a city. Our
aim is to get an insight about how the interaction between drivers with distinctive
behaviors can influence a 2D \cite{brunet1997} traffic flow. Specifically, we study a system where part of the agents respects a traffic rule, while others ignore it. It is a very common situation in many countries with poor traffic education, from which an imitation behavior, ``do as the others do,'' usually emerges. In such a context, we explore the effect that the introduction of a core of ``well educated'' drivers, abiding by the law, affects the collective behavior and the efficiency of the system in terms of, for example, speed.

In the vast literature on traffic there are few studies which use the tools of game theory. Yet, the behavior of drivers can easily and adequately be treated as a \emph{strategy} in the game theoretical sense. The framework we follow in the present work is reminiscent of other studies in the dynamics of pedestrians \cite{Fukui1999,Muramatsu1999,Takimoto2002}. In particular, Baek et al.~\cite{Baek2009} have analyzed a cellular automaton where agents move along a passageway in both directions.  When encountering other agents they take a step to the right with probability $p$ and to the left with probability $1-p$. In this way, they play a coordination game in which two strategies are considered: traffic rule abiders and traffic rule ignorers.

Other game theoretical approaches have been applied on networks of transportation approached
from an economic point of view \cite{Youn2008,Su2007}. Certainly, there are many
instances were the mathematics of cooperating and defecting strategies, of imitation
and education, can be applied in traffic systems. The education of traffic rules is
perhaps the epitome where the need of reaching a consensus of cooperation in the
system produces a global benefit. A similar problem in a more abstract setting has
been studied in \cite{laguna2010}, and we aim to applying some of those ideas here.

\section{Model}

\label{sec:1}With the purpose of obtaining an insight of the mechanisms operating behind the
emergent collective patterns, we analyze a simple model that contains enough
ingredients from the real systems. The model city is a square with straight streets
arranged in a regular lattice. For the sake of simplicity in the description,
imagine that the streets run either in the North-South direction (the
\emph{longitudinal} streets) or the East-West (the \emph{transversal} streets). All
streets are one-way, single-lane roads. The direction of traffic alternates in both
sets of streets, as is usual in many real world situations.

The vehicles are modelled as a cellular automaton. A car can occupy the space
between two intersections, and advance one block at each time step of the automaton
dynamics. Double occupancy of the blocks is not allowed: the cars can advance only
if the block ahead is empty. Interaction of cars occur only at the intersections, in
a manner that will be described below. We keep a constant number of vehicles by
allowing those drivers that exit the lattice from a border to reenter the system. To
prevent artifacts that a periodic boundary condition would produce in the rather
small systems we use to mimic the streets of a real world city, these cars are
randomly returned to empty slots at the border of any street, regardless the street
from which they came out.

The dynamics proceeds in discrete steps. At each step all the intersections of the
city are updated in a random order. The effect of this is an asynchronous update of
the positions of the vehicles, which prevents some artifacts that a sequential
update produces, especially at high density. A simulation run consists of the
repetition of this update step $nL^2$ times (with usually $n=100$), which provides a
reasonable number of interactions between drivers. We observed that the flow reaches
a stationary state after a short transient, which allowed us to take measurements
and time averages during the course of the simulation. The results shown below
correspond, furthermore, to ensemble averages of 100 such simulation runs.

There is only one traffic rule in the system: drivers shall give way to vehicles
approaching from the right at intersections. This rule is widely used in countries
with right-hand traffic, and applied at all intersections where it is not overridden
by priority signs or traffic signals, neither of which are present in our model.
Drivers are either \emph{cooperators} (who abide by the rule, and yield to drivers
coming from their right) or \emph{defectors} (who ignore the rule).

\begin{figure}[h!]
\resizebox{0.47\textwidth}{!}{%
  \includegraphics{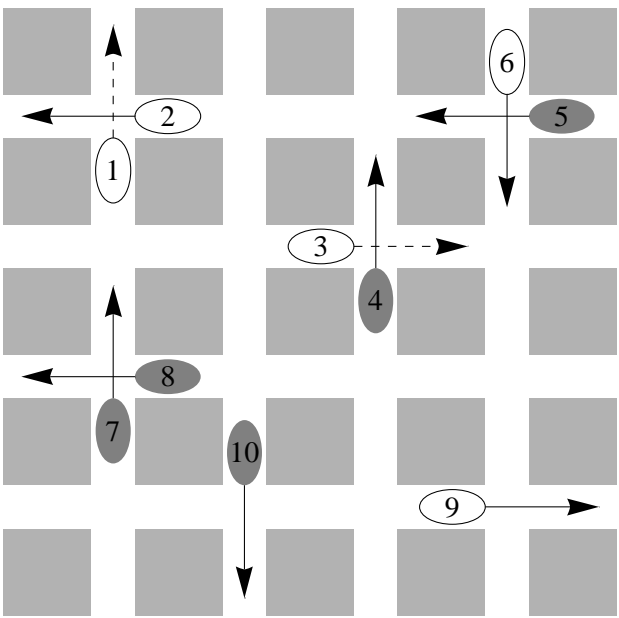}
}
\caption{Sketch of a distribution of vehicles in a model city. Cooperators and defectors are represented by white and grey ovals respectively. The arrows show
the  direction  of  motion of the cars. Dashed arrows indicate drivers that give way to the other;
solid arrows correspond to drivers that cross (or try to cross) the intersection. The four interactions shown (besides the ``free'' drivers 9 and 10) correspond to the four types of encounter of strategies that
can occur at intersections.}
\label{figure0}       
\end{figure}

When two vehicles approach an intersection at the same time there is an interaction
that determines the order of crossing and the time involved. There may be four
different situations according to the drivers' strategies, as follows. \emph{(i) Both
drivers are cooperators.} One of them gives way to the other, who has right of way
according to priority of the right. \emph{(ii) The driver coming from the right is a
defector, and the other is a cooperator.} This situation is analogous to (i),
since the cooperator yields to the defector, and the defector just ignores the rule.
\emph{(iii) The cooperator approaches from the right, and the other
driver is a defector.} This is the reverse of (ii). The defector does not yield, while the cooperator assumes his
right to cross and attempts to do it. In this case we suppose that some time is lost in the
fake pass. \emph{(iv) Both drivers are defectors.} In this situation we suppose a collision
that entails a longer loss of time than in case (iii).  The time that a driver needs to
cross the intersection can be used as a measure of the cost involved in the game
situation, analogous to the payoff of the usual formulation of formal games. The
loss of time that occurs in situations (iii) and (iv) can be a substantial contribution
to the cumulative cost, and can be interpreted as pass fakes, fines or even crashes.
The crosses actually happen only if the slot just ahead of the intersection is free.
Figure~\ref{figure0} shows a cartoon summarizing the description of the system.

The costs involved in rules (i) to (iv) are summarized in the following table, that
show the time in units of simulation steps. A \emph{left (right) driver} is a driver
approaching the intersection from the left (right) of the other driver. A driver
that yields loses a time step, incurring into a cost of $2$ for crossing. The
parameters $a$ and $b$ characterize the delay incurred into in case of more serious
interactions. Observe that, in the interaction (iii), we assume that both drivers lose the same time, $a$. This is a simplification of a situation that could be, of course, more complex in real life. For example, the cooperator might stop completely, the offending defector cross, and then the cooperator continue his way. This could be implemented in the present formulation with an additional parameter $c>a$ for the cooperator. A sensible choice, in the present cellular automaton formulation, would be $c=a+1$. The results do not change significantly in such a case, reflecting a robustness of the model with respect to details that might be difficult to quantify. On the other hand, one does not sensibly expect that $c>a+1$. Even if one would allow it, a thorough reformulation of the rules would be necessary to accommodate for the possibility of a vehicle standing still even when the interacting one has followed his way. In the present work we prefer to keep the model rules simple, as formulated.

\begin{center}
\begin{tabular}{|c|c|c|}
\hline
\backslashbox{Left driver}{Right driver} & C & D  \\
\hline
C & \backslashbox{2}{1} & \backslashbox{2}{1} \\
\hline
D & \backslashbox{a}{a} & \backslashbox{b}{b} \\
\hline
\end{tabular}
\end{center}


\section{Results}

Let us begin with the analysis of systems with quenched strategies. Each driver
receives, at the beginning, a strategy that remains unchanged during the course
of the simulation. Strategies are randomly drawn, uniformly with a probability $p_C$
for cooperation and its complement $1-p_C$ for defection. The density of cooperators
is then $p_C$, uniformly distributed in the city. The simulations runs we show below
consist of $100L^2$ time steps, in a square city of $L^2/2$ intersections, and with
a time step that updates, randomly and asynchronously, as many intersections. After
a transient of $L^2$ time steps we measure the instantaneous average velocity in the
system, defined as the ratio of moving vehicles to their total number, and compute
the time average of it, denoted  $\langle v\rangle$ below. Average velocity of
cooperators and of defectors ($\langle v_C\rangle$ and $\langle v_D\rangle$) are
defined similarly. Besides, we also measure the average velocity of each vehicle as
the ratio of the number of steps it makes during the simulation, to the total
duration. From this ensemble we compute the distributions $P(v_C)$ and $P(v_D)$,
characterizing the movement of the two classes of drivers.

\subsection{Density dependence}

\begin{figure}[h!]
\resizebox{0.45\textwidth}{!}{%
  \includegraphics{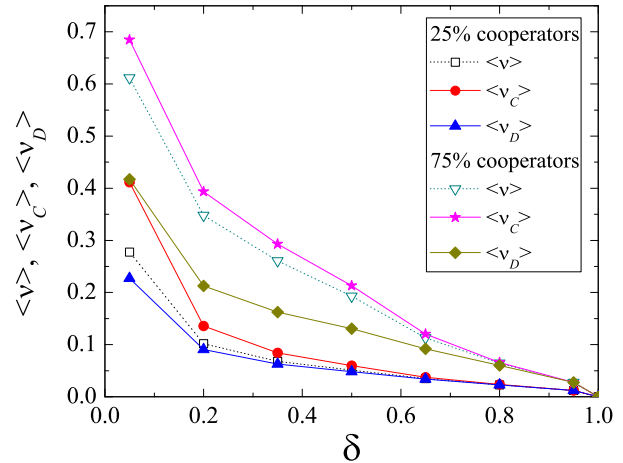}
} \caption{Mean velocity of the system, of cooperators and of defectors ($\langle
v\rangle$, $\langle v_C\rangle$ and $\langle v_D\rangle$ respectively) as a function
of the density of vehicles, $\delta$. We show two different cooperation scenarios,
as indicated in the legend: 25\% and 75\% of cooperators in the population. The
costs involved in the interactions are $a=3$, $b=100$. Data are averaged over 100
realizations.}
\label{v-dens}       
\end{figure}

Figure~\ref{v-dens} shows the average velocities as a function of vehicle density
$\delta$ in a $20\times 20$ square lattice. Two sets of curves are shown,
corresponding to different strategy compositions of the system. The lowest three
curves correspond to a system with 25\% cooperators. The upper three, on the other
hand, correspond to a very cooperative system with 75\% cooperators. These two
situations provide a good characterization of both cooperative and non-cooperative
ensembles.

As Fig. \ref{v-dens} shows, the velocity decreases monotonically with vehicle
density. This is so for the system average velocity $\langle v\rangle$ as well as
for the cooperators and the defectors. Observe, also, that the average velocity of
cooperators is greater than that of defectors at all densities. This reflects the
fact that cooperators are less prone to the higher costs involved in the violation
of the right hand rule ($b\gg a>2$). Moreover, observe also the effect of system
composition on the average velocity. When the system is more cooperative, defectors
also get a benefit. Indeed, the velocity of defectors is greater in a system with
75\% of cooperators than that of cooperators in a system with 25\% of
cooperators---at least for intermediate densities. We will return to these matters
later on.

\begin{figure}[h!]
\resizebox{0.45\textwidth}{!}{%
  \includegraphics{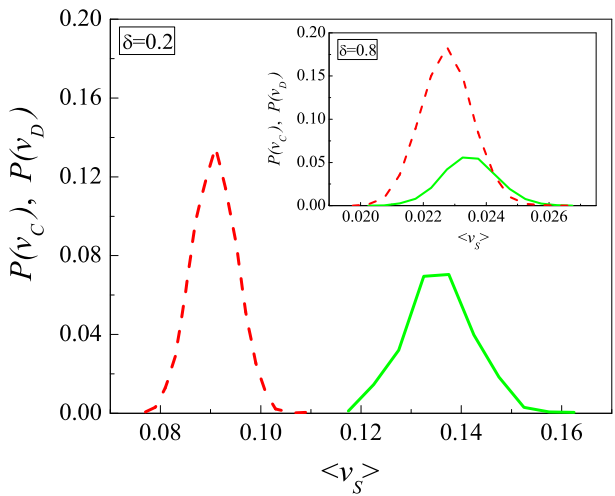}
} \caption{Distribution of the velocity of cooperators (full curve) and defectors
(dotted curve) for densities $\delta=0.2$ in the main panel and $\delta=0.8$ in the
insert. The costs are $a=3$, $b=100$. Data are averaged over 100 realizations.}
\label{histo-25}       
\end{figure}

We have also computed the distribution of time-averaged individual velocities, and
checked that the mean values shown in Fig. \ref{v-dens} are a good characterization
of them. Figure \ref{histo-25} shows a typical example. The distributions are bell
shaped around their mean values, with narrow dispersion. Only when the mean values
of cooperators and defectors are very close (as it happens when the density is high)
the distributions show some overlap as we see in the insert of the Figure
\ref{histo-25}.

\subsection{Imitation dynamics}

The results presented in the previous section provide a first step in our analysis
of the dynamics of cooperation and defection in the simple traffic model. We will
now introduce the key ingredient of strategy imitation. Let us suppose that there
are two idiosyncrasies in the drivers. Some of them adopt their driving strategy not
because of conviction or belief in the convenience of respecting the traffic rule,
but by imitation of other drivers. It is a common habit in many societies with
little enforcement of the rules and a low level of education of them: ``Do as the
other do.'' Besides these ``imitators,'' a core of drivers may well be \emph{bona
fide} cooperators, representing a small group of drivers that have been well
educated in the traffic rules, and are convinced of their application. These ``core
cooperators'' do not imitate the strategy of other drivers, and always cooperate.

We implement this imitation dynamics in the following way. Imitating drivers can
change their strategies based on the frequency of encounters with drivers that
display one or the other strategy. Since it is unreasonable that drivers change
strategy too often, the imitation takes place every $\tau$ simulation steps, based
on the frequency of interactions during the previous $\tau$ steps. The fraction of
encounters with either strategy define the individual probabilities of imitation:
\begin{align}
P_C &= \frac{f_C}{f_C+f_D}, \\
P_D &= 1-p_C,
\end{align}
where $f_s$ is the number of interactions with strategy $s$ during the previous
$\tau$ steps. Observe that the left driver participating in encounters of type (i) and (ii)---a cooperator---cannot infer the strategy of the right driver. They always give way, regardless of whether the right driver is a cooperator or a defector. In these situations, the cooperator adds $0.5$ both to $f_C$ as $f_D$. In any other type of encounter each driver is able to infer the strategy used by  the other driver, and will add $1$ to $f_C$ or to $f_D$ as appropriate. Based on the defined probabilities, drivers can change their strategy by imitating the majority, if they play the other strategy:
\begin{align}
C &\rightarrow D~~\text{with probability } P_D,\\
D &\rightarrow C~~\text{with probability } P_C.
\end{align}

It is reasonable to assume that drivers in a small town tend to interact with the
same drivers over and over again, while those in a big city do not. In this spirit
we have made $\tau$ depend on the size of the city, as $L^2$ simulation steps.

The numerical results show that the dynamics of imitation establishes a balance between the number of cooperators and defectors. Independently of the initial conditions, the system reaches an equilibrium in the composition of strategies. The average number of cooperators and defectors  at the end of the realizations, with the parameters we used, are $56\%$ and $44\%$ respectively. Moreover, these same values of equilibrium are observed even in extreme cases when a single cooperator or defector is introduced at the beginning of the simulations in populations in which all other drivers have the opposite strategy (besides fortuitous extinctions of one of the populations when it is very small, due to stochastic fluctuations).

This approach to equilibrium can be understood in a mean field approximation by analyzing a master equation for the probability densities. Consider the following equation for the probability density of cooperators:
\begin{equation}\label{PCevol}
\frac{d\rho_{C}}{dt}=W(C|D)\rho_D-W(D|C)\rho_C,
\end{equation}
where $W(S|S')$ is the transition probability for a driver with strategy $S'$ to switch to $S$, and $\rho_C$ and $\rho_D=1-\rho_C$ are the probability densities of cooperators and defectors respectively at time $t$.

Given the lack of discrimination of cooperators driving on the left at intersections, in a well mixed approximation their probability to switch to defection is $W(D|C)\approx 1/2$. On the other side, defectors build up a probability of switching to cooperation based on their encounters with cooperators, so we have $W(C|D)=\rho_C$. With these, Eq.~(\ref{PCevol}) becomes a logistic equation of the form:
\begin{equation}
\frac{d\rho_{C}}{dt}=\rho_C(1/2-\rho_C).
\label{logistic}
\end{equation}
The stable equilibrium solution of Eq.~(\ref{logistic}) is $\rho^*_C=1/2$, close to the observed stationary value. Incidentaly, observe that a \emph{partial} discrimination of the cooperators, of a fraction $\epsilon<1$ of the defectors in their encounters, can be taken into account with a transition probability of the form $W(D|C)=\epsilon\rho_D+(1-\epsilon)/2$. It is easy to show that the ensuing logistic equation also has $\rho^*_C=1/2$ as only attractor. The timescale of the evolution, however, becomes dependent on $\epsilon$ as $2(1-\epsilon)^{-1}$.

Besides the imitators, we set a fraction $f_{CC}$ of drivers to form a core of cooperators.
They respect the traffic rules strictly, and we intend to study their effect in the dynamics. After the system has reached a stationary state they are randomly chosen, set as
cooperators, and will not change their strategy for the remaining course of the run.
They represent a set of drivers who have been educated to abide by the law,
disregarding the behavior of the rest of the system. We are interested in the
reaction of the system as a whole, as a result of the imitation of strategies, and
its dependence on the size of this core of cooperation.

\begin{figure}[h!]
\resizebox{0.45\textwidth}{!}{%
  \includegraphics{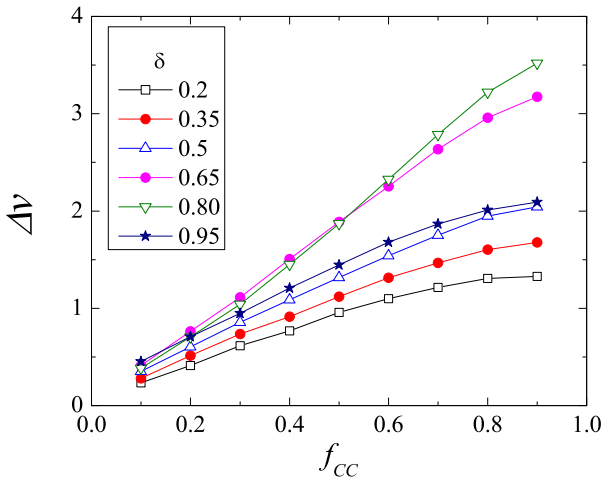}
} \caption{Coefficient $\Delta v$, measuring the increase of the system velocity after the
introduction of a core of cooperators, versus the relative size of the core in the
population. The curves show the behavior for different values of the vehicle density
$\delta$, as shown in the legend. The initial fraction of cooperators in the system
is 25\% in all cases. The cost parameters are $a=3$, $b=100$. Data are averaged over
100 realizations.}
\label{deltav}       
\end{figure}

As a global measure of the influence of the core we define a coefficient:
\begin{equation}
\Delta v = \frac{v_a-v_b}{v_b},
\end{equation}
where $v_b$ and $v_a$ are respectively the average system velocities calculated
before and after the establishment of the core. Figure~\ref{deltav} shows the effect
of the core size on $\Delta v$ for different vehicle densities. It is clear that the
existence of the core of cooperators represents a benefit for the system as a whole,
in terms of the velocity of the traffic flow.  Moreover, the growth of $\Delta v$
is monotonous, without any indication of a transition for some critical core size.
This  behavior is very general, and we have checked it for a wide range of model
parameters.

The meaning of this result may be important for the design of education plans. It implies
that not only it is necessary to educate and convince as many drivers as possible of
the importance of the respect of the rules, but also that the imitation of behavior
does not produce a collective critical turnover of a defective system.

Observe, also, that the actual value of the growth of performance may be sometimes dubious, since the growth of velocity can be marginal in some situations,
particularly for low and high densities of traffic. This is an indication that other
measures are necessary to achieve specific goals of traffic fluidity, besides the
education of a core of drivers.

\begin{figure}[h!]
\resizebox{0.45\textwidth}{!}{%
  \includegraphics{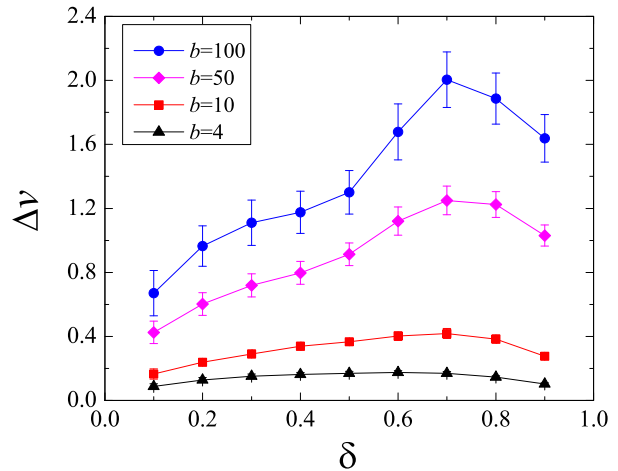}
} \caption{Coefficient $\Delta v$ as a function of vehicle density, for different
costs of the interaction between defectors. The other parameters are $a=3$ and
$f_{CC}=0.5$. Error bars represent the mean square deviation of the data in 100
realizations.}
\label{dv-vs-d}       
\end{figure}

As a final comment on Fig.~\ref{deltav}, observe that the increase of the velocity is not
monotonous in the density: the greatest values of $\Delta v$ correspond to
intermediate values of $\delta$. To show this dependence we plot, in Fig.
\ref{dv-vs-d}, $\Delta v$ as a function of $\delta$ for a single value of the core
size, $f_{CC}=0.5$. The plot shows that the increase of the average velocities of the system
are greater near and in excess of $0.5$ occupancy of the lattice. Moreover, greater
penalties on defector drivers entail a substantial improvement of these average
speeds in the system.

\section{Conclusions}

We studied a cellular automata model of vehicular traffic based on game theory. Our
simple model assumes that only two behaviors are possible: drivers either respect or
ignore the rule that gives priority to the vehicle driving on the right. The two
classical strategies of two-strategies formal games---cooperation and
defection---are immediately associated with those. In our framework, pairs of drivers
can interact at crossings where each vehicle pay a temporal cost according to its
strategy.

We studied first the average velocity achieved in systems with different and fixed
proportions of cooperators and defectors. We found that the velocity decreases
monotonically with density, driving away the possibility of a sharp transition at any traffic density.  Even for high vehicle
densities, the average speed of cooperators is greater than that of defectors,
independently of what strategy is a majority in the population. This seems to imply
that cooperators always benefit when abiding by the law. However, in a population
essentially cooperative, defectors can exploit this situation for their own benefit,
obtaining higher speeds than those achieved when the population is essentially
defective.

Additionally, we introduced a dynamic of imitation. Each driver---at a slower time scale than the update of the state of the system---can change their strategy with a probability calculated from
their perception of the strategy of those they meet at the crossroads. This imitation represents the attitude of drivers who have not received a proper education in traffic rules; instead, they try to adjust their behavior to that of the rest of the system. We have explored the effect of establishing a core of law-abiding drivers in this environment, in the form of a sub-population of cooperators that do not change their behavior during the dynamical evolution.

The existence of a core of cooperators represent a
benefit for the whole system, in the sense that higher speeds are achieved. This benefit is gradual and continuous in its dependence on the size of the core set. That is, the combination of the core together with the emulation
attitude provided by the dynamic of imitation, does not produce a critical turnover
of a defecting system. This is a point to take into consideration in the formulation of driver's education programs (even for drivers that already have a permit or license) or road safety campaigns. Remarkably, this
benefit reaches a maximum at intermediate densities of traffic, and it is greater when the
penalties impose higher costs. The results are very general in the simple model reported in the present work, which tries to capture a set of minimal important details of the actual system. They will be further explored with the addition of more complex rules and interactions in future investigations. Moreover, some of the dynamical features observed in the model could be considered as suggestions for field observations.


\begin{thebibliography}{99}
\bibitem{helbing2001} D. Helbing, Traffic and related self-driven many-particle systems, Rev. Mod. Phys. \textbf{73}, 1067-1141 (2001).

\bibitem{chowdhuryetal2000} D. Chowdhury, L. Santen and A. Schadschneider, Statistical physics of vehicular traffic and some related systems, Phys. Rep. \textbf{329}, 199-329 (2000).

\bibitem{lighthill1955} M. J. Lighthill and G. B. Whitham, On kinematic waves: II. A theory of traffic on long crowded roads, Proceedings of the Royal Society A \textbf{299}, 317-345 (1955).

\bibitem{richards1956} P. I. Richards, Shock waves on the highway, Opns. Res. \textbf{4}, 42-51 (1956).

\bibitem{nagel1992} K. Nagel and M. Schreckenberg, A cellular automaton model for freeway traffic, J. de Physique I (France) \textbf{2}, 2221-2229 (1992).

\bibitem{brunet1997} L. G. Brunnet and S. Gonçalves, Cellular automaton block model of traffic in a city, Physica A \textbf{237}, 59-66 (1997).

\bibitem{Fukui1999} M. Fukui and Y. Ishibashi, Jamming transition in cellular automaton models for pedestrians on passageway, J. Phys. Soc. Jpn. \textbf{68}, 3738 (1999).

\bibitem{Muramatsu1999} M. Muramatsu, T. Irie, and T. Nagatani, Jamming transition in pedestrian counter flow Physica A \textbf{267}, 487 (1999).

\bibitem{Takimoto2002} K. Takimoto, Y. Tajina, and T. Nagatani, Effect of partition line on jamming transition in pedestrian counter flow, Physica A \textbf{308}, 460 (2002).

\bibitem{Baek2009} S. K. Baek, P. Minnhagen, S. Bernhardsson, K. Choi, and B. J. Kim, Flow improvement caused by agents who ignore traffic rules, Phys. Rev. E {\bf 80}, 016111 (2009).

\bibitem{Youn2008} Youn, H., M. Gastner, and H. Jeong, Price of anarchy in transportation networks: Efficiency and optimality control, Phys. Rev. Lett. {\bf 101}, 128701 (2008).

\bibitem{Su2007} B.B. Su, H. Chang, Y.-Z. Chen and D.R. He, A game theory model of urban public traffic networks Physica A {\bf 379}, 291-297 (2007).

\bibitem{laguna2010} M.F. Laguna, G. Abramson, S. Risau-Gusman and J.R. Iglesias, Do the right thing, J. Stat. Mech. P03028 (2010).
\end{thebibliography}
\end{document}